\documentclass[conference]{IEEEtran}
\IEEEoverridecommandlockouts
\usepackage{cite}
\usepackage{amsmath,amssymb,amsfonts}
\usepackage{algorithmic}
\usepackage{multirow}
\usepackage{booktabs} 
\usepackage[inline, shortlabels]{enumitem}
\usepackage{subfig}
\usepackage{graphicx}
\usepackage{textcomp}
\usepackage{xcolor}
\def\BibTeX{{\rm B\kern-.05em{\sc i\kern-.025em b}\kern-.08em
    T\kern-.1667em\lower.7ex\hbox{E}\kern-.125emX}}

\def\TagAddition{A}
\def\TagConversion{C}
\def\TagSubtraction{S}

\def\MaxMicrostep{M\_M}
\def\MostRecentDNET{DN}
\def\LastSkippedNET{SN}
\def\MaxMicrostep{\ensuremath{M_{\text{max}}}}

\usepackage{amsmath}
\usepackage{graphicx}
\usepackage{xcolor}
\usepackage{ifthen}
\usepackage{comment}
\usepackage{enumitem} 
\usepackage[hidelinks]{hyperref}
\newcommand{\code}[1]{\textsf{\small#1}}
\newboolean{showcomments}
\ifdefined\DRAFT
  \setboolean{showcomments}{true} 
\else
  \setboolean{showcomments}{false}
\fi
\makeatletter
\newcommand{\mynote}[3]{%
  \ifthenelse{\boolean{showcomments}}{%
    {\textbf{\textcolor{#3}{(#1 $\triangleright$ #2)}}}}%
  {%

  }%
}

\newcommand{\Rmnum}[1]{\expandafter\@slowromancap\romannumeral #1@}
\makeatother
\usepackage[scaled=1.04]{couriers}
\definecolor{stringColor}{rgb}{0.1,0.5,0.1}
\usepackage{listings}
\lstdefinelanguage{LF}{
  keywords={msec, sec, timer, startup, shutdown, state, main, actor, handler, reaction, preamble, target, reactor, composite, trigger, input, output, constructor, new, action, clock, logical, physical, after, import, from, private, public, method, time, interleaved, extends},
  keywordstyle=\color{black}\bfseries,
  ndkeywords={class, export, boolean, throw, implements, this, int, if, else, float},
  ndkeywordstyle=\color{darkgray}\bfseries,
  identifierstyle=\color{black},
  sensitive=false,
  comment=[l]{//},
  morecomment=[s]{/*}{*/},
  commentstyle=\color{black}\ttfamily,
  stringstyle=\color{stringColor}\ttfamily,
  morestring=[b]',
  morestring=[b]"
}
\begin{document}


\title{%
	\vspace{-2em}
	\begin{center}
		\small This is an authors' copy of the paper  to appear in  Proceedings of the 22nd ACM-IEEE International 
		Conference on Formal Methods and Models for System Design (MEMOCODE'24)
	\end{center}
	\vspace{0.3em}
	Efficient Coordination for Distributed Discrete-Event Systems
}


\author{\IEEEauthorblockN{Byeonggil Jun}
\IEEEauthorblockA{\textit{Arizona State university}\\
byeonggil@asu.edu}
\and
\IEEEauthorblockN{Edward A. Lee}
\IEEEauthorblockA{\textit{University of California, Berkeley}\\
eal@berkeley.edu}
\and
\IEEEauthorblockN{Marten Lohstroh}
\IEEEauthorblockA{\textit{University of California, Berkeley}\\
marten@berkeley.edu}
\and
\IEEEauthorblockN{Hokeun Kim}
\IEEEauthorblockA{\textit{Arizona State university}\\
hokeun@asu.edu}
}
\IEEEaftertitletext{\vspace{-2.2\baselineskip}}

\maketitle
\begin{abstract}
Timing control while preserving determinism is often a key requirement for ensuring the safety and correctness of distributed cyber-physical systems (CPS).
Discrete-event (DE) systems provide a suitable model of computation (MoC) for time-sensitive distributed CPS.
The high-level architecture (HLA) is a useful tool for the distributed \emph{simulation} of DE systems,
but its techniques can be adapted for \emph{implementing} distributed CPS.
However, HLA incurs considerable overhead in network messages conveying timing information between the distributed nodes and the centralized run-time infrastructure (RTI).
This paper gives a novel approach and implementation that reduces such network messages while preserving DE semantics.
An evaluation of our runtime demonstrates that our approach significantly reduces the volume of messages for timing information in HLA.
\end{abstract}


\section{Introduction}\label{sec:introduction}


Distributed cyber-physical systems (CPS) interacting with the physical world and connected over the networks are becoming more pervasive and widely used.
A distributed CPS often requires
timing control over the network with determinism, ensuring the same outputs and behavior for given initial conditions and inputs.
One of the ways to support such determinism in distributed CPS is the high-level architecture (HLA)~\cite{IEEE:10:HLA}, an IEEE standard for discrete event (DE) system simulation.
The methods of HLA have also been extended to support features for system implementation (vs. simulation).

An HLA-based system includes a run-time infrastructure (RTI) such as CERTI~\cite{noulard2009certi}, a centralized coordinator, and federates, distributed individual nodes.
During the simulation or execution of an HLA-based system, federates react to events in a logical time order.
Events have a timestamp on the globally agreed logical timeline~\cite{LohstrohEtAl:23:LogicalTime}.
To ensure that federates see events in logical time order, federates exchange signals that include timing information with the RTI.
However, the frequent exchange of such signals drastically increases network overhead, potentially causing issues, as shown in our case study of implementing distributed CPS using HLA~\cite{JunEtAl:23:EventDetection}.

This paper introduces methods that significantly reduce the number of signals in HLA-based distributed systems.
We use an open-source coordination language, Lingua Franca (LF)~\cite{LohstrohEtAl:21:Toward}, as our baseline because LF provides an extended, working implementation with mechanisms similar to HLA.
The centralized coordinator of LF is loosely based on HLA~\cite{BateniEtAl:23:Risk}.
We examine the purpose of each signal that is used for coordinating logical time and find scenarios where signals are used inefficiently.
Then, we provide solutions to eliminate the signals that are not essential for ensuring determinism in HLA.


\section{Related Work}\label{sec:RelatedWork}



A number of algorithms and methods for synchronization mechanisms have been proposed for distributed discrete event (DDE) simulation~\cite{jafer2013synchronization}.
Among those, HLA~\cite{IEEE:10:HLA} has been widely used and standardized by the IEEE.

There has been research on optimizing the performance of DDE systems and HLA with RTI.
Rudie \textit{et al.}~\cite{rudie2003minimal} present a strategy to minimize the communication in DDE systems by producing minimal sets of communications.
Wang and Turner~\cite{wang2004optimistic} propose an optimistic time synchronization of HLA using an RTI with a rollback ability when events turn out to be incorrectly scheduled during the optimistic simulation,
although such approaches are not for deployment where a rollback is impossible.
COSSIM~\cite{tampouratzis2020novel} provides time synchronization that can trade off the timing accuracy and performance with relaxed synchronization for CPS simulation using an IEEE HLA-compliant interface.
Distinct from COSSIM, our approach improves efficiency while strictly synchronizing a DDE system with an enhancement to standard HLA.


\section{Background}\label{sec:Background}

Lingua Franca (LF) is an open-source coordination language and runtime implementing reactors~\cite{Lohstroh:EECS-2020-235, Lohstroh2020Reactors}.
Reactors adopt advantageous semantic features from established models of computation, particularly actors~\cite{Agha1997Actors}, logical execution time~\cite{Kirsch2012LET}, synchronous reactive languages~\cite{benveniste2003synchronous}, and discrete event systems~\cite{lee2014discrete}.
LF also enables deterministic interactions between physical and logical timelines~\cite{LohstrohEtAl:21:Toward}.

LF adopts the superdense model of time for logical time\cite{LohstrohEtAl:21:Toward}.
Each tag $g \in \mathbb{G}$ is a pair of a \emph{time value} $t \in \mathbb{T}$ and a \emph{microstep} $m \in \mathbb{N}$.
The time value $t$ includes limiting values, \emph{NEVER},  a time value earlier than any other, and \emph{FOREVER}, a time value larger than any other, represented by symbols $-\infty$ and $\infty$ in this paper, respectively.


Below are formal set definitions of $\mathbb{N}, \mathbb{T}$ and $\mathbb{G}$ for our discussion in this paper:
\begin{itemize}
	\item $\mathbb{N}$: A set of non-negative integers that can be represented by a 32-bit unsigned integer.
	\item $\mathbb{T}$: A set of non-negative integers that can be represented by a 64-bit signed integer $\cup~\{-\infty\}$.
	\item $\mathbb{G}$: A set defined as $\{(t, m)~|~t \in \mathbb{T}\setminus \{-\infty, \infty\},~m\in \mathbb{N} \}
	\cup \{(-\infty,0), (\infty, \MaxMicrostep)\}$.
\end{itemize}

The set of tags, $\mathbb{G}$, forms a totally ordered set where, given two tags $g_a = (t_a, m_a)$ and $g_b = (t_b, m_b)$, $g_a < g_b$ if and only if 
\begin {enumerate*} [1) ]%
\item $t_a<t_a$ or \item $t_a=t_a$ and $m_b<m_b$
\end {enumerate*}

We introduce a function $\TagAddition$, an operation like tag addition.
Here, we handle overflow in the formalization, recognizing that useful programs should not encounter overflow;
We formally define $\TagAddition\colon \mathbb{G} \times \mathbb{G} \longrightarrow\mathbb{G}$ as shown below:

\begin{equation}\label{eq:A}
	\small
	\begin{aligned}
		\TagAddition(g_a, g_b) = \TagAddition((t_a, m_a), (t_b, m_b)) = \\
		\begin{cases}
			(t_a,m_a + m_b), & \text{if } 0 \le t_a < \infty \land t_b =0 \land m_a + m_b < \MaxMicrostep \\ 
			(t_a + t_b, m_b), & \text{if } 0 \le t_a < \infty \land t_b > 0  \land t_a + t_b < \infty \\
			(t_a,\MaxMicrostep), & \text{if } 0 \le t_a < \infty  \land t_b =0  \land m_a + m_b \ge \MaxMicrostep \\
			(\infty,~\MaxMicrostep), & \text{if } 0 \le t_a < \infty  \land t_b > 0  \land t_a + t_b \ge \infty \\
			(-\infty,~0), & \text{if } t_a = -\infty \lor t_b -\infty\\
			(\infty,~\MaxMicrostep) & \text{if } t_a = \infty \\
\end{cases}
\end{aligned}
\end{equation}
\noindent
Note that if $t_b$ is greater than 0, $\TagAddition((t_a, m_a), (t_b, m_b))$ yields a tag that has $m_b$ as the microstep, effectively ignoring $m_a$.

For simplicity, from now on, we will use an \emph{elapsed} tag, the elapsed time and microstep since the startup of the program, instead of an \emph{actual} tag with the time from January 1, 1970.
Formally, when an actual tag is $g = (t, m)$ and the startup time of the program $t_{s}$, then the elapsed tag is $(t - t_{s}, m)$.


\begin{figure}
	\centering
	\includegraphics[width=0.75\columnwidth]{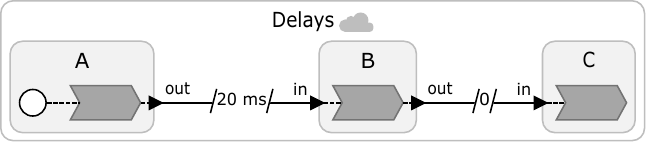}
	\caption{An example LF program with delays.}
	\label{fig:Delays}
\end{figure}

\begin{table}
\centering
\begin{tabular}{llll}
	Name       & Payload & Description                   & Direction          \\
	\toprule
	MSG$_{ij}$   & Tag \& Message & Tagged Message          & $j$ to $i$ via RTI         \\
	LTC$_j$    & Tag & Latest Tag Complete          & $j$ to RTI         \\
	NET$_j$    & Tag & Next Event Tag                & $j$ to RTI         \\
	TAG$_i$    & Tag & Tag Advance Grant             & RTI to $i$         \\
	\bottomrule
\end{tabular}
\caption{Types of messages and signals related to time management that are exchanged by federates and the RTI.\label{tab:message_types}}
\end{table}

\figurename~\ref{fig:Delays} shows a diagram of an LF program named \emph{Delays} with three reactor instances, $a$, $b$, and $c$, which are instances of reactor classes $A$, $B$, and $C$.
The cloud symbol indicates that this is a federated program where each top-level reactor becomes a federate that can be deployed to remote machines.
Dark gray chevrons indicate reactions that are executed when triggered.
The white circle in $a$ denotes a \code{startup} trigger.

LF supports the concept of \textit{logical delay} to explicitly represent logical time elapsing through a connection.
As \code{after} delays in LF are specified using time values, to make use of function $\TagAddition$, we need to convert the delay represented by a time value to a tag.
We introduce a function $\TagConversion\colon \mathbb{T} \longrightarrow \mathbb{G}$ for converting a time value of delay to a tag:  

\begin{equation}\label{eq:C}
	\small
	\begin{aligned}
		\TagConversion(d) = 
		\begin{cases}
			(0,~1), & \text{if } d = 0 \\
			(d,~0), & \text{if  } 0 < d < \infty \\
			(0,~0), & \text{if } d = -\infty \\
			(\infty,~\MaxMicrostep) & \text{if } d = \infty \\
		\end{cases}
	\end{aligned}
\end{equation}
\noindent
Specifically, when a reactor sends a message with tag  $g_s = (t_s, m_s)$ through a connection with logical delay $d \in \mathbb{T}$, the resulting tag $g_d$ is $\TagAddition(g_s, \TagConversion(d))$.
Let $D_{ij}$ be the \emph{minimum tag} increment delay over all connections from $j$ to $i$.
This means that	if a federate $j$ sends a message with a tag $g_j$, this may cause a message for $i$ with a tag $\TagAddition(g_j, D_{ij})$, but no earlier.

A federate can advance its logical time to a tag $g$ only if the RTI guarantees that the federate will not later receive any messages with tags earlier than or equal to $g$.
The RTI and federates continuously exchange the signals LTC, NET, and TAG that are briefly described in \tablename~\ref{tab:message_types} to keep track of each federate's state and manage time advancement.
We use a function $G$ to denote the payload tag of a signal or a message.
For example, $G(\text{MSG}_{ij})$ is the tag of the message $\text{MSG}_{ij}$.
When federate $i$ receives $\text{MSG}_{ij}$, it schedules an event at $G(\text{MSG}_{ij})$ to process the message.

LTC$_j$ (Latest Tag Complete) is sent from a federate $j$ to the RTI to notify that federate $j$ has finished $G(\text{LTC}_j)$.

NET$_j$ (Next Event Tag) is sent from federate $j$ to the RTI to report the tag of the earliest unprocessed event of $j$.
This signal promises that $j$ will not later produce any messages with tags earlier than $G(\text{NET}_j)$ unless it receives a new message from the network with a tag earlier than $G(\text{NET}_j)$.

TAG$_i$ (Tag Advance Grant) is sent by the RTI to federate $i$.
When $i$ receives this signal, it knows it has received every message with a tag less than or equal to $G(\text{TAG}_i)$.
It can now advance its tag to $G(\text{TAG}_i)$ and process all events with tags earlier than or equal to $G(\text{TAG}_i)$.
If a federate $i$ has no upstream federate, then $i$ can advance its tag without TAG.

For each federate $j$, the RTI maintains variables $N_j$ and $L_j$ and a priority queue $Q_j$ called the \textit{in-transit message queue} to predict $j$'s future behavior.
$N_j$ is the tag of the latest received NET$_j$.
$L_j$ is the tag of the latest received LTC$_j$.
$Q_j$ is a priority queue that stores tags of in-flight messages that have been sent to $j$, sorted by tag.
When the RTI forwards a message to $j$, it stores the tag in $Q_j$, and when the RTI receives LTC$_j$, it removes tags earlier than or equal to $G(\text{LTC}_j)$ from $Q_j$.
Let $H(Q_j)$ denote the head of $Q_j$.

When the RTI receives NET$_i$,
it updates $N_i$ and decides whether to send TAG$_i$.
Let $U_i$ denote the set of federates immediately upstream of $i$ (those with direct connections).
Let $g = \min\limits_{j \in U_i} D(L_j, D_{ij})$.
If $g \ge N_i$, then the RTI grants TAG$_i$ with a tag $g$, allowing it to process its events.
If $g < N_i$, it may still be possible to grant a TAG$_i$ by computing $B_i$, the \textbf{earliest (future) incoming message tag} for node $i$.
The RTI can send TAG$_i$ with a tag $\min(N_i, H(Q_i))$ if $B_i > \min(N_i, H(Q_i))$.

To calculate $B_i$, consider an immediate upstream federate $j \in U_i$.
The earliest possible tag of a future message from $j$  that the RTI might see is  $\min (B_j, N_j, H(Q_j))$.
Consequently, we can compute $B_i$ recursively as:
\begin{equation}\label{eq:EIMT}
B_i = \min_{j \in U_i} (A(\min (B_j, N_j, H(Q_j)), D_{ij})))
\end{equation}
The RTI can easily calculate this quantity unless there is a cycle (paths from $i$ back to itself) with no \code{after} delays~\cite{donovan2024strongly}.
In this paper, we simply ignore federates within zero-delay cycles and do not apply our optimizations to them.
\section{Inefficiency In Time Management Protocol}\label{sec:Ineff}
In this section, we discuss inefficiencies in LF's HLA-based timing coordination. 
As described in Section~\ref{sec:Background}, a federate sends NET every time it completes a tag, which is unnecessary.

Federate $i$ sends NET$_i$ for two reasons: (1) to notify $i$'s next event tag to the RTI to let the RTI compute TAG signals for $i$'s downstream federates and (2) to request a TAG$_i$ to advance $i$'s tag to $G(\text{NET}_i)$.
Therefore, a NET$_i$ signal is unnecessary
if it results in no TAG signal for downstream federates and if $i$ can safely advance to a tag $G(\text{NET}_i)$.

\begin{figure}
	\centering
	\includegraphics[width=0.5\columnwidth]{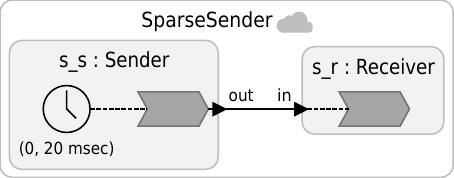}
	\caption{An LF program where the federate Sender sends messages sparsely.}
	\label{fig:SparseSender}
	\vspace{-15pt}
\end{figure}

We show how unnecessary NET signals are produced using a simple LF example shown in \figurename~\ref{fig:SparseSender}.
The timer in the upstream federate $s\_s$ triggers $s\_s$'s reaction every 20 ms.
Assume $s\_s$ is a federate polling a distance sensor and the downstream federate $s\_r$ processes the sensing result.
For instance, the sensor can be used for detecting an emergency situation in an autonomous vehicle or detecting cars entering and exiting a parking lot~\cite{JunEtAl:23:EventDetection}.
The sensor only occasionally detects an interesting event, so it sends a message sparsely.

\begin{figure}
	\centering
	\includegraphics[width=0.58\columnwidth]{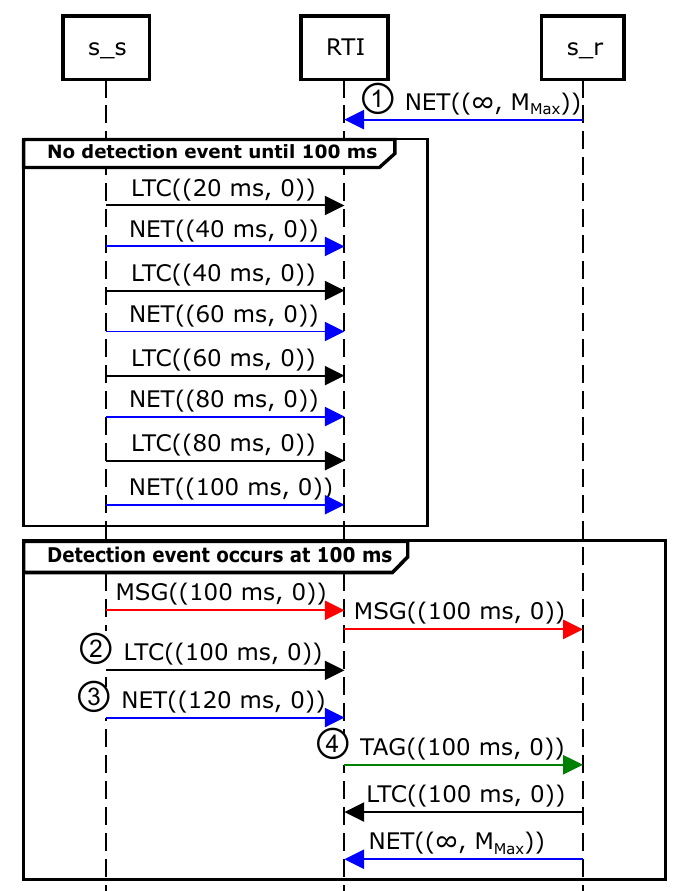}
	\caption{An execution trace of the program in \figurename~\ref{fig:SparseSender}.}
	\label{fig:SparseSenderTraceBaseline}
	\vspace{-15pt}
\end{figure}

\figurename~\ref{fig:SparseSenderTraceBaseline} shows the trace of the first 100 ms of execution of \figurename~\ref{fig:SparseSender}.
The trace does not show the signals at startup tag $(0,~0)$ that are irrelevant to our discussion.
Federate $s\_s$ sends a tagged message every 100~ms in this program.
After the startup tag, federate $s\_r$ sends NET$_{s\_r}((\infty,~\MaxMicrostep))$ in \textcircled{1}, which indicates that it has no event to execute, while $s\_s$ sends NET signals every 20~ms.
Most of the NET signals from $s\_s$ are unnecessary because $s\_s$ can advance its tag without TAG signals and the RTI cannot grant TAG$_{s\_r}((\infty,~\MaxMicrostep))$ to $s\_r$ with those NET signals.
At $(100~ms,~0)$, $s\_s$ sends a tagged message via the RTI and the RTI knows that $\min(N_{s\_r}, H(Q_{s\_r}))$ is $(100~ms,~0)$.
So the RTI sends TAG$_{s\_r}((100~ms,~0))$ in \textcircled{4} based on LTC$_{s\_s}((100~ms,~0))$ and NET$_{s\_r}((120~ms,~0))$, signals \textcircled{2} and \textcircled{3}, respectively.
Thus, $\text{NET}((120~ms,~0))$ is necessary.


\section{Downstream Next Event Tag}\label{sec:DNET}
We introduce a new signal, Downstream Next Event Tag (DNET), to eliminate unnecessary NET signals.
The RTI sends DNET$_j$ to federate $j$ with the tag $G(\text{DNET}_j)$ when all NET$_j$ signals such that $G(\text{NET}_j) \le G(\text{DNET}_j)$ are not necessary for computing TAG for $j$'s downstream federates.
Then, $j$ does not have to send NET$_j$ when $j$'s next event tag is earlier than or equal to $G(\text{DNET}_j)$, and $j$ can advance its logical time.
Note that $j$ still must send NET$_j$ signals in case $j$ cannot advance to its next event's tag.



The RTI has to calculate $G(\text{DNET}_i)$ carefully to prevent $i$ from skipping sending necessary NET$_i$ while removing every unnecessary NET$_i$ signal.
If $G(\text{DNET}_i)$ is too early, $i$ would send unnecessary NET$_i$,
and if $G(\text{DNET}_i)$ is too late, $i$ may not send a necessary NET$_i$, resulting in $i$'s downstream federates being unable to advance their tags.

%

Let $\bar{D_j}$ denote the set of federates ``transitive'' downstream of federate $j$, specifically, those that are connected from $j$ via one or more chained connections.
For every downstream node $i$, the RTI tries to grant TAG$_i$ of $\min(N_i, H(Q_i))$.
Thus, the RTI has to find an upper bound of the tags of unnecessary NET$_j$ signals.
Let $g$ be the latest tag to which $j$ can advance without sending NET$_j$ for a federate $i \in \bar{D_j}$.
$g$ must be the latest tag that satisfies the condition: 
\[
\small
\TagAddition(g, D_{ij}) <= \min(N_i, D_{ij})
\]
For example, if $\min(N_i, H(Q_i))$ is $(2~s,~3)$ and $D_{ij}$ is $(0,~3)$, then a tag $(2~s,~0)$ is an upper bound of unnecessary NET$_j$'s tag.
Concretely, NET$_j$ with $G(\text{NET}_j) = (2~s,~0)$ is unnecessary because $B_i =A((2~s,~0), (0,~3)) = (2~s,~3)$ is not later than $\min(N_i, H(Q_i))$.
If, on the other hand, the RTI receives NET$_j$ with $G(\text{NET}_j) = (2~s,~1)$, the earliest tag among tags later than $(2~s,~0)$, it \textit{can} grant a TAG$_i$ $((2~s,~3))$ because $B_i = A((2~s,~1), (0,~3)) = (2~s, 4) > (2~s,~3)$.


To calculate the upper bound of the tags of unnecessary NET$_j$ signals, we define a function $\TagSubtraction$ that behaves like tag subtraction.
What we really need is subtraction, but because A saturates the microstep on overflow, there is no function S such that if $g = \TagSubtraction(g_a, g_b)$ then $\TagAddition(g, g_b) = \TagAddition(\TagSubtraction(g_a, g_b), g_b) = g_a$, which is what a true subtraction function would do.
Instead, we define function $\TagSubtraction$ that returns a tag $g = \TagSubtraction(g_a, g_b)$ where $g$ is the latest tag of the set of tags that satisfy:

\begin{equation}\label{eq:Cond}
	\small
	\TagAddition(g, g_b) = \TagAddition(\TagSubtraction(g_a, g_b), g_b) <= g_a
\end{equation}
Formally, we define the function $\TagSubtraction\colon \mathbb{G} \times \mathbb{G}_b \longrightarrow\mathbb{G}$ as:
\begin{equation}\label{eq:S}
	\small
	\begin{aligned}
		\TagSubtraction(g_a, g_b) = 
		\TagSubtraction((t_a, m_a), (t_b, m_b)) = \\
		\begin{cases}
			(-\infty,~0), & \text{if } g_a = -\infty \lor g_a < g_b \\
			(t_a - t_b, m_a - m_b), & \text{if } \infty > t_a \ge t_b = 0 \land m_a \ge m_b \\
			(t_a - t_b, \MaxMicrostep), & \text{if } \infty > t_a \ge t_b > 0 \land m_a \ge m_b \\
			(t_a - t_b - 1, \MaxMicrostep), & \text{if } \infty > t_a > t_b > 0 \land m_a < m_b \\
			(\infty,~\MaxMicrostep) & \text{if } t_a = \infty \\
			
		\end{cases}
	\end{aligned}
\end{equation}
where $\mathbb{G}_b$ is $\mathbb{G} \setminus \{(-\infty,~0),~(\infty,~\MaxMicrostep)\}$.
When we use function $\TagSubtraction$ to compute $G(\text{DNET}_j)$, $g_a$ is $\min(N_i,H(Q_i))$ and $g_b$ is $D_{ij}$.
If $i \in \bar{D_j}$, the value $D_{ij}$ cannot be $(\infty,~\MaxMicrostep)$ because $D_{ij}$ of $(\infty,~\MaxMicrostep)$ means there is no path from $j$ to $i$.
Also, $D_{ij}$ cannot be $(-\infty,~0)$ because $D_{ij}$ is always greater than or equal to $(0,~0)$ (``no delay'' is encoded to $(0,~0)$).
We know that the RTI \textit{cannot} send TAG$_i$ with a tag $\min(N_i,H(Q_i))$ if $j$ sends NET$_j$ such that $G(\text{NET}_j) \le \TagSubtraction(\min(N_i,H(Q_i)),D_{ij})$.
Thus, we compute $G(\text{DNET}_j)$, the upper bound of the tags of NET$_j$ signals that are unnecessary for $j$'s \emph{every} downstream federate, as:
\begin{equation}
	\small
	\min\limits_{\forall i \in \bar{D_j}} (\TagSubtraction(\min(N_i, H(Q_i)), D_{ij}))
	\label{eq:g_dnet_calc}
\end{equation}
When the value of Equation~(\ref{eq:g_dnet_calc}) changes due to any update to $N_i$ or $H(Q_i)$ and $j$ has not sent any necessary NET$_j$ signals,
the RTI needs to send new DNET$_j$ to $j$.

Now we describe how federates deal with DNET signals.
Each federate $j$ maintains a variable $\MostRecentDNET_j$ which stores the most recent DNET$_j$'s tag.
At the start of the execution, $j$ initializes $\MostRecentDNET_j$ to $(-\infty,~0)$.
Upon deciding not to send its next event tag, $j$ stores the tag in a variable $\LastSkippedNET_j$,
which denotes the last skipped next event tag.
When $j$ sends NET$_j$, it resets $\LastSkippedNET_j$ to $(-\infty,~0)$.
The variable $\LastSkippedNET_j$ is needed when a skipped NET signal is revealed to be necessary later.

There are three cases where a federate $j$ must update or use $\MostRecentDNET_j$ or $\LastSkippedNET_j$ to decide whether to send NET$_j$ signals.

First, when $j$ completes a tag $g_j$ and has a next event at a tag $g'_j$, $j$ compares $\MostRecentDNET_j$ against $g'_j$.
Assume that $j$ can advance to $g'_j$.
\begin {enumerate*} [1) ]%
\item  If $\MostRecentDNET_j < g'_j$, the RTI needs NET$_j$ ($g'_j$) for at least one of $j$'s downstream federates. $j$ must send NET$_j (g'_j)$ and reset $\LastSkippedNET_j$ to $(-\infty,~0)$.
\item If $\MostRecentDNET_j \ge g'_j$, none of federate in $\bar{D_j}$ requires NET$_j (g'_j)$. The federate $j$ decides not to send NET$_j(g'_j)$. And thus, $j$ stores the tag $g'_j$ to $\LastSkippedNET_j$.
\end {enumerate*}

Second, when $j$ receives a new DNET$_j$, it checks whether $G(\text{DNET}_j)$ is earlier than $\LastSkippedNET_j$.
\begin {enumerate*} [1) ]%
\item  If $G(\text{DNET}_j) < \LastSkippedNET_j$, at least one downstream federate is waiting for $j$'s NET$_j$ with a tag later than $G(\text{DNET}_j)$.
So $j$ must send NET$_j$ with the tag $\LastSkippedNET_j$ and stores $G(\text{DNET}_j)$ in $\MostRecentDNET_j$ and resets $\LastSkippedNET_j$ to $(-\infty,~0)$.
\item If $G(\text{DNET}_j) \ge \LastSkippedNET_j$, no federate requires NET$_j (\LastSkippedNET_j)$. $j$ only changes $\MostRecentDNET_j$ to the new $G(\text{DNET}_j)$.
\end {enumerate*}

Third, when $j$ sends a new tagged message $\text{MSG}_{ij}$ to federate $i$ with a destination tag $g_t$, $j$ compares $g_t$ against $DN_j$.
If
\begin {enumerate*} [1) ]%
\item  If $g_t < \MostRecentDNET_j$, $j$ knows that $i$ has an event at $g_t$. 
Thus, $j$ updates $\MostRecentDNET_j$ to $g_t$.
This allows $j$ to send necessary NET$_j$ without having to wait for a new DNET$_j$.
\item If $g_t \ge \MostRecentDNET_j$, no action is required. $j$ will send NET$_j$ after it completes the current tag anyway.
\end {enumerate*}

\begin{figure}
	\centering
	\includegraphics[width=0.6\columnwidth]{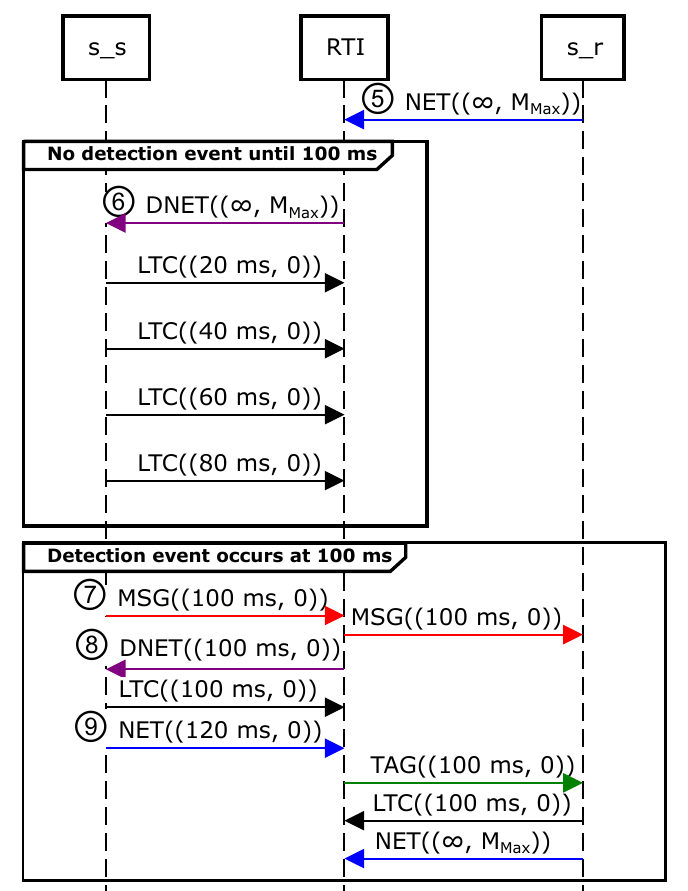}
	\caption{An execution trace of the program in \figurename~\ref{fig:SparseSender} with DNET.}
	\label{fig:SparseSenderTraceWithDNET}
\end{figure}

\figurename~\ref{fig:SparseSenderTraceWithDNET} shows the trace of the LF program in \figurename~\ref{fig:SparseSender} where DNET signals are used for removing unnecessary NET signals.
When the RTI receives NET$_{s\_r}((1~s,~0))$ in \textcircled{5} from $s\_r$, it sends DNET$_{s\_s}((1~s,~0))$ in \textcircled{6} to $s\_s$ as $\TagSubtraction((1~s,~0), (0,~0))$ is $(1~s,~0)$.
Consequently, $s\_s$ does not send NET signals until $(100~ms,~0)$.
At $(100~ms,~0)$, $s\_s$ sends $\text{MSG}_{s\_rs\_s}((100~ms,~0))$ in \textcircled{7}.
Now, the RTI knows that $s\_r$ has an event at $(100~ms,~0)$ as $\min(N_{s\_r}, H(Q_{s\_r}))$ is $(100~ms,~0)$.
Thus, the RTI sends DNET$_{s\_s} (100~ms,~0)$ in \textcircled{8}.
Upon receiving the new DNET$_{s\_s}$, $s\_s$ sends NET$_{s\_s}((120~ms,~0))$ in \textcircled{9}, as $(120~ms,~0) > (100~ms,~0)$.

\section{Evaluation}\label{sec:EVAL}

We evaluate our approach in comparison with the baseline implementation of LF.
We take the example in \figurename~\ref{fig:SparseSender} as a microbenchmark.
We simulate the examples with our approach and the baseline and compare the number of NET signals.

We assume the actual event detection happens every 5 seconds while varying the sensing periods (the timer periods).
Note that, in practice, the actual detection usually occurs more sparsely.
For example, in a real smart factory or a distance sensing system on a vehicle, a defective product or an object (hopefully) does not appear at intervals of a few seconds.
%

\begin{table}
	\centering
	\begin{tabular}{|l|r|r|r|r|r|}
			\hline
			\textbf{Timer Period} & \textbf{5 ms} & \textbf{10 ms} & \textbf{20 ms} & \textbf{50 ms} & \textbf{100 ms} \\
			\hline
			Baseline  & 100,161 & 50,191 & 25,193 & 10,195 & 5,195 \\
			\hline
			Our Solution  & 677 & 385 & 301 & 288 & 297 \\
			\hline
		\end{tabular}
	\caption{Number of exchanged NET signals during the 500 seconds of runtime with timer periods from 5 ms to 100 ms, using the SparseSender example shown in   \figurename~\ref{fig:SparseSender}.}
	\label{tab:NumSignals}
\end{table}

\tablename~\ref{tab:NumSignals} shows the count of the NET signal.
Our solution remarkably reduces the number of signals for every example.
The number of signals decreases by at most 148 times.
We observe that our solution becomes more effective as the sparsity grows (as the timer period decreases while the event detection period is constant).

\section{Conclusion}\label{sec:CON}
In this paper, we propose an efficient timing coordination for DDE systems.
Our evaluation using an extended version of an open-source DDE system shows the proposed solution significantly reduces the cost of exchanging network signals.
We note that the effectiveness of the proposed approach varies depending on the topology and sparsity of the application.


\bibliographystyle{IEEEtran}
\bibliography{Refs}

\begin{thebibliography}{10}
\providecommand{\url}[1]{#1}
\csname url@samestyle\endcsname
\providecommand{\newblock}{\relax}
\providecommand{\bibinfo}[2]{#2}
\providecommand{\BIBentrySTDinterwordspacing}{\spaceskip=0pt\relax}
\providecommand{\BIBentryALTinterwordstretchfactor}{4}
\providecommand{\BIBentryALTinterwordspacing}{\spaceskip=\fontdimen2\font plus
\BIBentryALTinterwordstretchfactor\fontdimen3\font minus
  \fontdimen4\font\relax}
\providecommand{\BIBforeignlanguage}[2]{{%
\expandafter\ifx\csname l@#1\endcsname\relax
\typeout{** WARNING: IEEEtran.bst: No hyphenation pattern has been}%
\typeout{** loaded for the language `#1'. Using the pattern for}%
\typeout{** the default language instead.}%
\else
\language=\csname l@#1\endcsname
\fi
#2}}
\providecommand{\BIBdecl}{\relax}
\BIBdecl

\bibitem{IEEE:10:HLA}
IEEE, ``{IEEE} standard for modeling and simulation ({M\&S}) high level
  architecture ({HLA})-- framework and rules,'' \emph{IEEE Std 1516-2010
  (Revision of IEEE Std 1516-2000) - Redline}, pp. 1--38, 2010.

\bibitem{noulard2009certi}
E.~Noulard, J.-Y. Rousselot, and P.~Siron, ``{CERTI}, an open source {RTI}, why
  and how,'' 2009.

\bibitem{LohstrohEtAl:23:LogicalTime}
M.~Lohstroh, E.~A. Lee, S.~Edwards, and D.~Broman, ``Logical time for reactive
  software,'' in \emph{Workshop on Timing-Centric Reactive Software (TCRS), in
  Cyber-Physical Systems and Internet of Things Week (CPSIoT)}.\hskip 1em plus
  0.5em minus 0.4em\relax ACM, 2023, Conference Proceedings.

\bibitem{JunEtAl:23:EventDetection}
\BIBentryALTinterwordspacing
B.-G. Jun, D.~Kim, M.~Lohstroh, and H.~Kim, ``Reliable event detection using
  time-synchronized iot platforms,'' in \emph{Proceedings of Cyber-Physical
  Systems and Internet of Things Week 2023}, ser. CPS-IoT Week '23.\hskip 1em
  plus 0.5em minus 0.4em\relax New York, NY, USA: Association for Computing
  Machinery, 2023, p. 355–360. [Online]. Available:
  \url{https://doi.org/10.1145/3576914.3587501}
\BIBentrySTDinterwordspacing

\bibitem{LohstrohEtAl:21:Toward}
\BIBentryALTinterwordspacing
M.~Lohstroh, C.~Menard, S.~Bateni, and E.~A. Lee, ``Toward a lingua franca for
  deterministic concurrent systems,'' \emph{ACM Trans. Embed. Comput. Syst.},
  vol.~20, no.~4, May 2021. [Online]. Available:
  \url{https://doi.org/10.1145/3448128}
\BIBentrySTDinterwordspacing

\bibitem{BateniEtAl:23:Risk}
S.~Bateni, M.~Lohstroh, H.~S. Wong, H.~Kim, S.~Lin, C.~Menard, and E.~A. Lee,
  ``Risk and mitigation of nondeterminism in distributed cyber-physical
  systems,'' in \emph{ACM-IEEE International Conference on Formal Methods and
  Models for System Design (MEMOCODE)}, 2023, Conference Proceedings.

\bibitem{jafer2013synchronization}
S.~Jafer, Q.~Liu, and G.~Wainer, ``Synchronization methods in parallel and
  distributed discrete-event simulation,'' \emph{Simulation Modelling Practice
  and Theory}, vol.~30, pp. 54--73, 2013.

\bibitem{rudie2003minimal}
K.~Rudie, S.~Lafortune, and F.~Lin, ``Minimal communication in a distributed
  discrete-event system,'' \emph{IEEE transactions on automatic control},
  vol.~48, no.~6, pp. 957--975, 2003.

\bibitem{wang2004optimistic}
X.~Wang, S.~J. Turner, M.~Y.~H. Low, and B.~P. Gan, ``Optimistic
  synchronization in {HLA} based distributed simulation,'' in \emph{Proceedings
  of the eighteenth workshop on Parallel and distributed simulation}, 2004, pp.
  123--130.

\bibitem{tampouratzis2020novel}
N.~Tampouratzis, I.~Papaefstathiou, A.~Nikitakis, A.~Brokalakis,
  S.~Andrianakis, A.~Dollas, M.~Marcon, and E.~Plebani, ``A novel, highly
  integrated simulator for parallel and distributed systems,'' \emph{ACM
  Transactions on Architecture and Code Optimization (TACO)}, vol.~17, no.~1,
  pp. 1--28, 2020.

\bibitem{Lohstroh:EECS-2020-235}
\BIBentryALTinterwordspacing
M.~Lohstroh, ``Reactors: A deterministic model of concurrent computation for
  reactive systems,'' Ph.D. dissertation, EECS Department, University of
  California, Berkeley, Dec 2020. [Online]. Available:
  \url{http://www2.eecs.berkeley.edu/Pubs/TechRpts/2020/EECS-2020-235.html}
\BIBentrySTDinterwordspacing

\bibitem{Lohstroh2020Reactors}
M.~Lohstroh, {\'I}.~{\'I}. Romeo, A.~Goens, P.~Derler, J.~Castrillon, E.~A.
  Lee, e.~R. Sangiovanni-Vincentelli, Alberto", M.~Edin~Grimheden, and W.~Taha,
  ``Reactors: A deterministic model for composable reactive systems,'' in
  \emph{Cyber Physical Systems. Model-Based Design}.\hskip 1em plus 0.5em minus
  0.4em\relax Cham: Springer International Publishing, 2020, pp. 59--85.

\bibitem{Agha1997Actors}
\BIBentryALTinterwordspacing
G.~A. Agha, \emph{Abstracting Interaction Patterns: A Programming Paradigm for
  Open Distributed Systems}.\hskip 1em plus 0.5em minus 0.4em\relax Boston, MA:
  Springer US, 1997, pp. 135--153. [Online]. Available:
  \url{https://doi.org/10.1007/978-0-387-35082-0_10}
\BIBentrySTDinterwordspacing

\bibitem{Kirsch2012LET}
\BIBentryALTinterwordspacing
C.~M. Kirsch and A.~Sokolova, \emph{The Logical Execution Time Paradigm}.\hskip
  1em plus 0.5em minus 0.4em\relax Berlin, Heidelberg: Springer Berlin
  Heidelberg, 2012, pp. 103--120. [Online]. Available:
  \url{https://doi.org/10.1007/978-3-642-24349-3_5}
\BIBentrySTDinterwordspacing

\bibitem{benveniste2003synchronous}
A.~Benveniste, P.~Caspi, S.~A. Edwards, N.~Halbwachs, P.~Le~Guernic, and
  R.~De~Simone, ``The synchronous languages 12 years later,'' \emph{Proceedings
  of the IEEE}, vol.~91, no.~1, pp. 64--83, 2003.

\bibitem{lee2014discrete}
E.~A. Lee, J.~Liu, L.~Muliadi, H.~Zheng, and C.~Ptolemaeus, ``Discrete-event
  models,'' \emph{System Design, Modeling, and Simulation using Ptolemy II},
  2014.

\bibitem{donovan2024strongly}
P.~Donovan, E.~Jellum, B.~Jun, H.~Kim, E.~A. Lee, S.~Lin, M.~Lohstroh, and
  A.~Rengarajan, ``Strongly-consistent distributed discrete-event systems,''
  \emph{arXiv preprint arXiv:2405.12117}, 2024.

\end{thebibliography}

\end{document}